\documentclass[apj,a4paper,12pt,useAMS]{emulateapj} %===== HERE for astro-ph (2nd choice)
\usepackage{amsmath} %==================================== HERE for astro-ph (2nd choice)

\usepackage{apjfonts} %============= here for astro-ph: (goes with `aastex' class)

\def\bum{$\beta$-UMi}
\newcommand\T{\rule{0pt}{2.6ex}}
\newcommand\B{\rule[-1.2ex]{0pt}{0pt}}

\slugcomment{The Astrophysical Journal, 665:55--66, 2007 August 10}

\shorttitle{MAXIPOL: Data Analysis and Results}
\shortauthors{Wu {\it et al.}}

\begin{document}

\title{MAXIPOL: Data Analysis and Results}

\author{J. H. P. Wu\altaffilmark{1}, J. Zuntz\altaffilmark{2}, 
M. E. Abroe\altaffilmark{3}, P. A. R. Ade\altaffilmark{4},
J. Bock\altaffilmark{5}, J. Borrill\altaffilmark{6,7}, J. Collins\altaffilmark{8}, \\
S. Hanany\altaffilmark{3}, A. H. Jaffe\altaffilmark{2}, B. R. Johnson\altaffilmark{9}, 
T. Jones\altaffilmark{3}, A. T. Lee\altaffilmark{8,10,7},  T. Matsumura\altaffilmark{3}, \\
B. Rabii\altaffilmark{8}, T. Renbarger\altaffilmark{3}, P. L. Richards\altaffilmark{8},
G. F. Smoot\altaffilmark{8,10,7}, R. Stompor\altaffilmark{11,6}, H. T. Tran\altaffilmark{8,7}, \\
C. D. Winant\altaffilmark{8}}

\altaffiltext{1}{Department of Physics, Institute of Astrophysics, \& Center for Theoretical Sciences,
								 National Taiwan University, Taipei 10617, Taiwan}
\altaffiltext{2}{Astrophysics Group, Blackett Lab, Imperial College, London, UK, SW7 2AZ}
\altaffiltext{3}{School of Physics and Astronomy, University of Minnesota, Minneapolis, MN, 55455, USA}
\altaffiltext{4}{School of Physics and Astronomy, Cardiff University, Cardiff, UK, CF24 3YB}
\altaffiltext{5}{Jet Propulsion Laboratory, Pasadena, CA, 91109, USA}
\altaffiltext{6}{Computational Research Division, Lawrence Berkeley National Lab, Berkeley, CA, 94720, USA}
\altaffiltext{7}{Space Sciences Laboratory, University of California, Berkeley, CA, 94720, USA}
\altaffiltext{8}{Department of Physics, University of California, Berkeley, CA, 94720, USA}
\altaffiltext{9}{Astrophysics, University of Oxford, Oxford, UK, OX1 3RH}
\altaffiltext{10}{Physics Division, Lawrence Berkeley National Lab, Berkeley, CA, 94720, USA}
\altaffiltext{11}{Laboratoire AstroParticule et Cosmologie, Universit{\'e} Paris -7, Paris, France}

%%%%%%%%%%%%%%%%%%%%%%%%%%%%%%%%%%%%%%%%%%%%%%%%%%%
%%%%%%%%%%%%%%%%%%%%%%%%%%%%%%%%%%%%%%%%%%%%%%%%%%%
%%%%%%%%%%%%%%%%%%%%%%%%%%%%%%%%%%%%%%%%%%%%%%%%%%%
\begin{abstract}
We present results from and the analysis of data from MAXIPOL, 
a balloon-borne experiment designed to measure the polarization 
in the Cosmic Microwave Background (CMB). 
MAXIPOL is the first CMB experiment to obtain results
using a rotating half-wave plate as a rapid polarization modulator. 
We report results from observations of a sky area of 8 deg$^2$ with 10-arcmin resolution,
providing information up to $\ell \sim 700$. 
We use a maximum-likelihood method to estimate maps of the $Q$ and $U$ Stokes parameters
from the demodulated time streams, 
and then both Bayesian and frequentist approaches to compute the $EE$, $EB$, and $BB$ power spectra.
Detailed formalisms of the analyses are given.  
We give results for the amplitude of the power spectra 
assuming different shape functions within the $\ell$ bins, 
with and without a prior $C_\ell^{EB}=C_\ell^{BB}=0$, and
with and without inclusion of calibration uncertainty. 
We show results from systematic tests 
including differencing of maps, analyzing sky areas of different sizes, 
assessing the influence of leakage from temperature to polarization, 
and quantifying the Gaussianity of the maps. 
We find no evidence for systematic errors.
The Bayesian analysis gives weak evidence for an $EE$ signal.  
The $EE$ power is $55^{+51}_{-45}~\mu\mbox{K}^2$ at the 68\% confidence level for $\ell=151$--$693$.
Its likelihood function is asymmetric and skewed positive 
such that with a uniform prior the probability of a positive $EE$ power is 96\%. 
The powers of $EB$ and $BB$ signals at the 68\% confidence level 
are $18^{+27}_{-34}~\mu\mbox{K}^2$ and $-31^{+31}_{-19}~\mu\mbox{K}^2$ respectively 
and thus consistent with zero.  
The upper limit of the $BB$-mode at the 95\% confidence level is $9.5~\mu\mbox{K}$. 
Results from the frequentist approach are in agreement within statistical errors.
These results are consistent with the current concordance $\Lambda$CDM model.
\end{abstract}


%%%%%%%%%%%%%%%%%%%%%%%%%%%%%%%%%%%%%%%%%%%%%%%%%%%
%%%%%%%%%%%%%%%%%%%%%%%%%%%%%%%%%%%%%%%%%%%%%%%%%%%
%%%%%%%%%%%%%%%%%%%%%%%%%%%%%%%%%%%%%%%%%%%%%%%%%%%
\section{Introduction}

Observations of the Cosmic Microwave Background (CMB)
have dramatically enhanced our understanding of the universe.
The recent focus has been on the detection of polarization in the CMB
because it provides information complementary to what can be learned from the temperature anisotropy.
The discovery and characterization of the polarization
not only confirms the cosmological interpretation of the origin of the temperature anisotropy and large-scale structures, 
but also improves the accuracy with which we measure parameters in our cosmological model,
such as the epoch of reionization.
So far detection of CMB polarization has been reported by
DASI \citep{dasi-05}, CBI \citep{cbi-04, cbi-05}, CAPMAP \citep{capmap-04},
BOOMERANG \citep{B03}, and WMAP \citep{wmap-06pol}.
Here we report results from MAXIPOL with particular emphasis 
on the data analysis procedure and cosmological results.
A companion paper \citep{maxipol-inst} emphasizes the polarimetric instrumentation and observations.

MAXIPOL flew from the NASA Columbia Scientific Ballooning Facility
in Ft.\ Sumner, New Mexico in May 2003.
A region of about 8 square degrees, with Galactic
coordinates $l$ between $110.69^\circ$ and $114.98^\circ$ and $b$
between $38.75^\circ$ and $42.49^\circ$, was scanned during 
a 7.6 hour night scan. 
This region was located near the star Beta Ursae Minoris (\bum).
The beam size was 10 arcminutes.  
We present data collected with 12 polarimeters
that have a center frequency of 140~GHz. 
Polarimetry was 
implemented by rotating a half-wave plate (HWP) at a frequency 
of 1.86~Hz and analyzing the modulating polarization with 
a stationary grid.
We refer the reader
to the companion paper \citep{maxipol-inst} and to 
other publications for a more thorough review 
of MAXIPOL and its predecessor MAXIMA
\citep{ma-hanany,ma-lee,ma-stompor,ma-jaffe,ma-wu-ng,ma-abroe}.

The characteristics of CMB radiation can be described using the four Stokes parameters: 
the intensity $I$, the linear polarization $Q$ and $U$, and the circular polarization $V$. 
The anisotropy in $I$ (also called the temperature $T$) has been well measured.
The circular component $V$ can only arise from parity violating physics and is believed to be absent from the CMB (though this has not been experimentally verified).
The $Q$ and $U$ parameters arise during the last-scattering process and have been the focus of recent experimental and theoretical work.  Their presence is evidence for the standard scenario of the last scattering process and their characteristics carry information complementary to the temperature anisotropy.
Alternative to the parameters $Q$ and $U$, 
one may express the polarization in terms of $E$ and $B$,
which are curl and divergence free polarization tensors respectively.
For noise-free all-sky data $Q$ and $U$ can be converted to and from $E$ and $B$ exactly.
Otherwise one can only make a statistical conversion between the two.
Models of cosmological evolution usually predict the CMB in the form of power spectra:
the auto-correlations $TT$, $EE$, and $BB$, and
the cross-correlations $TE$, $TB$, and $EB$.
In this paper we report on measurements of the $Q$ and $U$ Stokes parameters, 
and the corresponding $EE$, $EB$ and $BB$ polarization power spectra.

We used two alternative statistical approaches to extract
quantities of interest from the data,
the Bayesian and frequentist approaches.
Both have been used successfully in the analysis of cosmological data.
For a given set of data, the Bayesian approach gives the smallest error interval
for the estimated quantities, 
but its application is sometimes computationally intractable.  

In our analysis we demodulated the timestreams 
and subsequently used a Bayesian maximum-likelihood approach
to make best-fit maps of the Stokes parameters.  
We then estimated the $EE$, $BB$ and
$EB$ polarization power spectra of the CMB using both 
Bayesian and frequentist approaches.  
In both cases we accounted for known
instrumental effects and  conducted  tests for systematic errors.

This paper is organized as follows.
In Section~\ref{formalism} we describe our analysis formalism and procedures 
for estimating $Q$ and $U$ maps from the time-ordered data (TOD), and power spectra from these maps.
In Section~\ref{results}, we present our results, including the maps, the power spectra,
and systematic tests.
Our conclusions are given in Section~\ref{conclusion}.

%%%%%%%%%%%%%%%%%%%%%%%%%%%%%%%%%%%%%%%%%%%%%%%%%%%
%%%%%%%%%%%%%%%%%%%%%%%%%%%%%%%%%%%%%%%%%%%%%%%%%%%
%%%%%%%%%%%%%%%%%%%%%%%%%%%%%%%%%%%%%%%%%%%%%%%%%%%
\section{Formalism for Data Processing}
\label{formalism}

%%%%%%%%%%%%%%%%%%%%%%%%%%%%%%%%%%%%%%%%%%%%%%%%%%%
%%%%%%%%%%%%%%%%%%%%%%%%%%%%%%%%%%%%%%%%%%%%%%%%%%% time-domain
\subsection{Time-domain processing}
\label{lock-in}

The time-ordered data (TOD) were flagged for the presence of 
transient signals and calibrated using laboratory data and observations of Jupiter.
They were cut into segments separated by gaps longer than 30 seconds,
depending on both the flagging of transient signals and the stationarity of noise.
Various tests were performed to ensure the Gaussianity and stationarity of the noise
within each segment \citep{collins-phd}.
From each segment we estimated and removed an instrumental signal that was synchronous with the rotation of the HWP,
which we call HWP synchronous signal \citep{johnson-phd}, 
and deconvolved the instrumental filters.
See \citet{maxipol-inst} for more details about these data processing steps.

The TOD of each of the polarimeters can be modelled as
\begin{mathletters}
\begin{equation}
	d_t = {s_t^T} + {\epsilon} \left[ - s_t^Q \cos\phi_t + s_t^U \sin\phi_t \right] + n_t,
\end{equation}
where
\begin{equation}
	\phi_t = 4\beta_t - 2\alpha_t ,
\end{equation}
$d_t \equiv d(t)$, $t$ is time, 
$s_t^T \equiv T({\bf x}_t)$, $s_t^Q \equiv Q({\bf x}_t)$, $s_t^U \equiv U({\bf x}_t)$, 
$n_t$ is the instrumental noise at time $t$,
${\bf x}_t \equiv {\bf x}(t)$ is the sky position of the pointing at time $t$, 
and $\epsilon $ is the modulation efficiency of the polarimeter.
The units of $d_t$ are $\mu$K and thus a calibration factor converting from the 
measured voltage to temperature has already been included. 
We present our results in the WMAP
convention with the Stokes parameters $I$, $Q$, and $U$, 
taking the North Galactic Pole as the direction of reference for the polarization \citep{IAU}.
The angle $\alpha_t \equiv \alpha({\bf x}_t)$ is the rotation angle of a vector 
pointing along a great circle to the zenith, 
measured relative to the polarization reference vector on the sky.
The transmission axis of the polarization analyzer is oriented at 90 degrees to the zenith direction.
The angle $\beta_t \equiv \beta ({\bf x}_t)$
is the rotation angle of the HWP relative to the transmission axis of the 
polarization analyzer.
During the observations, $\alpha_t$ changed at a rate of 15 degrees per hour
giving a frequency of $f_\alpha \approx 1.16 \times 10^{-5}$~Hz
while $\beta_t$ varied at $f_\beta = 1.86$~Hz.
Thus $f_\phi \approx 4 f_\beta = 7.44$~Hz.
The temporal data $d_t$ are sampled at intervals of $\Delta t=4.8\times 10^{-3}$ seconds.
\end{mathletters}

\begin{figure}%[ht] 
\plotone{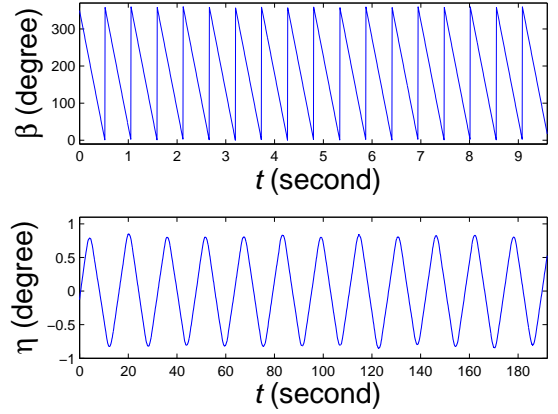}
	\caption
	{\small HWP rotation angle $\beta$ (upper panel) and
	scan angle $\eta$ (lower panel) from a typical 
	subset of the data.
	}
  \label{beta_eta}
\end{figure}

The telescope tracked the guide star {\bum} 
while scanning in azimuth by 2 degrees peak to peak      
at a constant frequency $f_\eta \simeq 0.06$~Hz for the majority of the data.
Here the subscript $\eta$ denotes the scan angle.
Figure~\ref{beta_eta} shows typical measurements of $\beta$ and $\eta$ 
from a subset of the data.

%--------------------------------------------------------FIG 1: beta, eta========here

\begin{figure}%[ht] 
\plotone{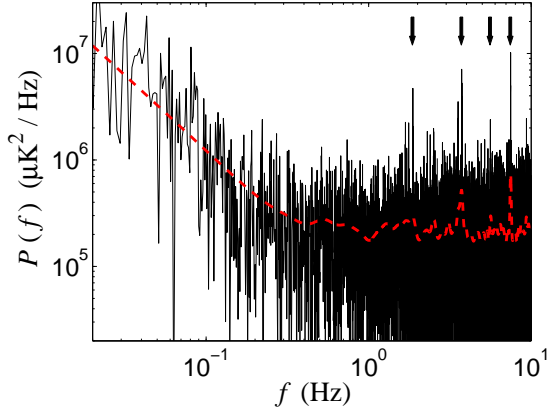}
  \caption
  {\small Power spectrum for a subset of the TOD $d_t$,
  after removal of the HWP synchronous signal (and after deconvolution of instrumental filters).
	The solid line is the raw spectrum, and the dashed line is its smoothed estimate.
	The arrows indicate multiples of the rotation frequency of the HWP $f_\beta$
	showing residuals of the HWP synchronous signal.
	We give a more complete discussion of this synchronous signal in \citet{maxipol-inst}.
}
  \label{Pf-dt}
\end{figure}

\begin{figure}%[ht] 
\plotone{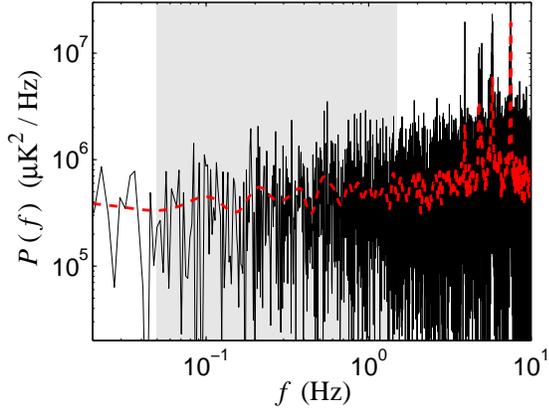}
  \caption
  {\small Power spectrum of the demodulated $Q$ of the data in Figure~\ref{Pf-dt}
  before the band-pass filtering; this is the quantity within the brackets of equation (\ref{lockin-a}).
	The solid line is the raw spectrum, the dashed line is its smoothed estimate,
	and the shaded area indicates the pass band of the filter. 
	The spectrum is consistent with white noise within this pass band.
  }
  \label{Pf-lockin}
\end{figure}

To obtain the Time-Ordered Polarization Data (TOPD)
we demodulated the TOD to produce independent data streams for $Q$ and $U$.
Because of the combination of the HWP rotation and the sky scan,
the CMB signal is in side-bands of the fourth harmonic of $\beta_t$.
Multiplication of the appropriate sinusoid and applying a band-pass filter
gives the TOPD from the TOD
\begin{mathletters}
\label{lockin}
\begin{eqnarray}
	d_t^Q & = &  \left\langle \frac{-2 d_t}{\epsilon} \cos \phi_t \right\rangle, \label{lockin-a}\\
	d_t^U & = &  \left\langle \frac{ 2 d_t}{\epsilon}  \sin \phi_t \right\rangle. \label{lockin-b}
\end{eqnarray}
The brackets denote a band-pass top-hat filter between 0.05~Hz and 1.5~Hz.
This filter selects the frequencies where signals are expected 
and removes residuals of the HWP synchronous signal, if they exist.
Figure~\ref{Pf-dt} shows a power spectrum of a section of the data $d_t$ before demodulation. 
The residual peaks (marked by arrows) indicate residuals of subtraction of the HWP synchronous signal.
The residuals are at harmonics of $f_{\beta}$. 
(Figure 6 in the companion paper~\cite{maxipol-inst} shows a similar power spectrum
for a different section of the data where the subtraction is more
complete.)
Figure~\ref{Pf-lockin} shows the power spectrum of the data $d_t$ 
multiplied by $-2 \cos \phi_t/\epsilon$, 
but before the band-pass filtering; 
that is, the quantity inside the brackets of equation~(\ref{lockin-a}). 
The gray area indicates the signal band selected by the band-pass filter.  
The low side of the band-pass is determined by the scan speed and 
the high side by the $\sim$1.3~Hz cut-off due to the beam.  
The power spectrum within the band is consistent with white noise and
residuals of the HWP synchronous signal are out of band. 
Simulations with different plausible power spectra for the underlying signal show that 
the band-pass filtering reduced the RMS of the $Q$ and $U$ maps by 4\% 
regardless of the details of the spectra.  
This loss of signal was compensated for by a Monte-Carlo approach, which we describe later.
\end{mathletters}

%-------------------------------- FIG2: power spectra of the "raw" (d_t)====here

%------------------------------- FIG 3: power spectra of the "locked-in" Q ==here

Several effects can bias our estimation of the TOPD when using equations (\ref{lockin}).
The beam convolution inevitably removes signal on angular scales smaller than $\sim 10$~arcminutes.
We rectified this by a deconvolution procedure during CMB power spectrum estimation
using recipes given in \citet{asym-beam}. These recipes also cope with the asymmetry in the beams.
This deconvolution is discussed in section~\ref{ps-estimation}.
A second effect is that 
an imperfect implementation of the demodulation may introduce a bias
in the estimation of the CMB signal. 
For example, the band-pass filtering is equivalent to a convolution in the time domain 
so that it induces correlations in the scan direction while giving rise to some loss of signal.
A third effect is that
the deglitching of the data for the transients creates small gaps in the TOD, 
and thus may have influenced the demodulation process.
The second and third effects were estimated and corrected by a Monte-Carlo approach,
as described in section \ref{ps-estimation}.

%%%%%%%%%%%%%%%%%%%%%%%%%%%%%%%%%%%%%%%%%%%%%%%%%%%
%%%%%%%%%%%%%%%%%%%%%%%%%%%%%%%%%%%%%%%%%%%%%%%%%%% map-making
\subsection{Map making}
\label{map-making}

We employed a standard maximum-likelihood method to obtain maps 
of $Q$ and $U$ from the TOPD. 
In the time domain the TOPD can be modelled as
\begin{equation}
  \label{d_t}
  d_t^X=s_t^X + n_t^X,
\end{equation}
where
$X=Q$ or $U$, $s_t^X$ is the CMB signal and $n_t^X$ is the instrumental noise.
We note that
$n_t^Q$ and $n_t^U$ are independent and thus uncorrelated
because the demodulation processes to obtain $d_t^Q$ and $d_t^U$ employ orthogonal kernels
(see Eqs.~(\ref{lockin})).
We model the CMB signal as
\begin{equation}
  \label{s_t}
  s_t^X=A_{tp}s_p^X,
\end{equation}
and use the Einstein summation convention when appropriate.
Here $A_{tp}$ is the pointing matrix giving the weight of pixel
$p$ in observation $t$, and $s_p^X$ is the CMB signal in the pixel.
We took the pointing operator $A_{tp}$ to be unity when
observing pixel $p$ at time $t$ and zero otherwise.
That is, we assumed the signal $s_t^X$ to be constant within pixel $p$. 
This model of pixelization induces an extra convolution effect
in addition to that from the beam.
To deal with these convolution effects we followed the recipes in \citet{asym-beam}.
These provide a way to transfer all these convolution effects into a single
$B_\ell$ in multipole space, 
which can then be deconvolved when estimating the CMB power spectrum $C_\ell$.

With this modeling,
we can estimate the pixelized maps $m_p^X$ from the temporal data $d_t^X$.
In the pixel domain we can also model $m_p^X$ as a linear sum of the signal and the noise components:
\begin{equation}
  m_p^X = s_p^X + n_p^X,
  \label{mp2}
\end{equation}
where
$n_p^X$ is the noise in the pixel domain.
Under the assumptions
that the noise in the temporal domain is Gaussian and
that all CMB maps are a-priori equally likely,
the maps $m_p^X$ can be estimated by maximizing the likelihood of the signal given the data.
This gives
\begin{equation}
  m_p^X = N_{pp'}^XA_{p't} ({N_{tt'}^X})^{-1} d_{t'}^X,
  \label{mp}
\end{equation}
where
$A_{p't}=A_{tp'}^{\rm T}$,
$N_{tt'}^X = \langle n_t^X {n_{t'}^X}^{\rm T}\rangle$ is
the time-time noise correlation matrix
and $N_{pp'}^X= \langle n_p^X {n_{p'}^X}^{\rm T}\rangle$ is the estimated pixel-pixel noise correlation matrix
given by
\begin{equation}
  N_{pp'}^X =\left[A_{pt} ({N_{tt'}^X})^{-1} A_{t'p'}\right]^{-1}.
  \label{Npp}
\end{equation}

We apply equations (\ref{mp}) and (\ref{Npp}) to $d_t^Q$ and $d_t^U$ to give the maps $m^Q_p$ and $m^U_p$,
respectively,
as well as the noise correlation matrices $N_{pp'}^Q$ and $N_{pp'}^U$.
The numerical implementation of these equations follows the method described in \citet{stompor-ma-map-01}.
We use square pixels that are 3 arcminutes on a side. 
To simplify notation we construct a column vector
\begin{equation}
m_q = \left( 
		\begin{array}{c}
			m^Q_p \\
			m^U_p
		\end{array}
	\right).
\end{equation}
Note that we use the subscript $p$ to denote pixels in the original maps 
and the subscript $q$ to denote pixels in the simplified notation.
Similarly we write
\begin{equation}
N_{qq'} = \left( 
		\begin{array}{cc}
			N^Q_{pp'} &  {\bf 0} \\
			{\bf 0} & N^U_{pp'}
		\end{array}
	\right).
\end{equation}
Here the off-diagonal blocks are zero in theory
because of the orthogonality between $n_t^Q$ and $n_t^U$.

\begin{figure}%[ht] 
\plotone{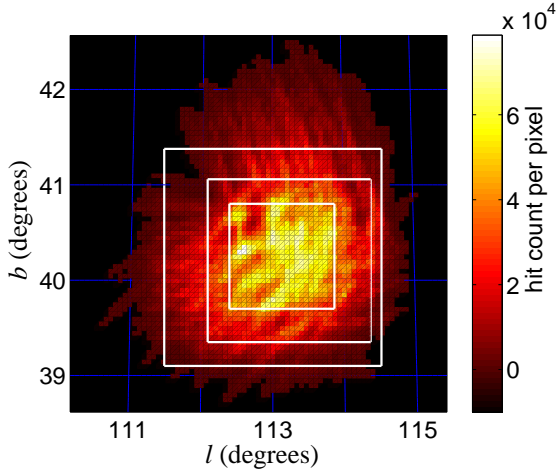}
  \caption
  {\small The sky coverage of MAXIPOL,
	presented as the number of 4.8-ms hits per 3-arcminute pixel.
	The squares are regions of 1.1, 1.7, and 2.3 degrees across,
	which we use to estimate the power spectra.
	Pixels with zero hit count but within the square regions have zero weighting
	in the process of power-spectrum estimation.}
  \label{fig-scan}
\end{figure}

After the maps are formed we apply a filtering to them 
\begin{equation}
	\label{weighting}
	m_q^{W} = W_{qq'} m_{q'},
\end{equation}
where $W_{qq'}$ is a filtering matrix. 
The choice of the filter depends on the subsequent step in the data analysis. 
For Bayesian power spectrum estimation the maps are left un-filtered 
(or, equivalently, we apply an identity filter). 
For the frequentist approach we apply a noise-weighting filter 
to cope with the anisotropic noise in the pixel domain.  
Because the hit count per pixel decreases towards the edge of the sky patch, 
as can be seen in Figure~\ref{fig-scan}, the S/N ratio is higher at the center.
To prevent the resulting power spectra from being dominated by the low-S/N pixels,
we use 
\begin{equation}
	\label{noise-weighting}
	W^{\rm N} = \left(U^T N^{-1} U\right)^{-1}U^TN^{-1},
\end{equation}
where $U$ is a column-vector with all entries equal to unity.
We also tried other forms of the filter used by other authors (e.g.\ \cite{B03}), 
and found that
as long as the biasing effect of such filtering is accounted for with the Monte-Carlo process
that we will describe later, the final CMB power spectra remain unchanged to a 10\% level.
For map display purposes we use a Wiener filter
\begin{equation}
	\label{wiener-filtering}
	W^{\rm WF} = S \left(  S + N \right)^{-1},
\end{equation}
where $S\equiv S_{qq'}=\langle s_q {s_{q'}}^{\rm T} \rangle$
is the signal-signal correlation matrix.
Because the Wiener filter amplifies expected signals in a model-dependent way it induces more bias than other filters that rely purely on the noise level measured from the data.
We thus do not use the Wiener filter for power spectrum estimation.

%----------------------------------------------------FIG 4: sky coverage ===here

%%%%%%%%%%%%%%%%%%%%%%%%%%%%%%%%%%%%%%%%%%%%%%%%%%%
%%%%%%%%%%%%%%%%%%%%%%%%%%%%%%%%%%%%%%%%%%%%%%%%%%%% ps-estimation
\subsection{Power-Spectrum Estimation}
\label{ps-estimation}

The $Q$ and $U$ can be expanded into spin-2 spherical harmonics
in the conventional way:
\begin{equation}
	\label{QU-lm}
	(Q\pm iU)({\bf n}) = 
	\sum_{\ell m} \left( a_{\ell m}^E \pm i a_{\ell m}^B \right) \, {_{\pm 2}}Y_{\ell m} ({\bf n}),
\end{equation}
where $a_{\ell m}^E$ and $a_{\ell m}^B$ are the coefficients for the $E$- and $B$-mode polarization, respectively,
and ${\bf n}$ is a unit vector directed in the direction of observation.
The polarization power spectra can then be defined as
\begin{equation}
	\label{Cl-lm}
	C_\ell^{YY'} = \frac{1}{2\ell+1}\sum_m a_{\ell m}^Y a_{\ell m}^{Y'*},
\end{equation}
where $Y$ and $Y'$ are either $E$ or $B$.
To estimate $C_\ell^{YY'}$, we use two approaches, one Bayesian and one frequentist, which are described in Sec.~\ref{ps-estimation-bayesian} and \ref{ps-estimation-frequentist} respectively.
In the following we first lay out the formalism which is general to both approaches.

Because the sky coverage of our observation is finite,
we do not probe independent $C_\ell^{YY'}$ for each multipole $\ell$.
Instead, we bin the $\ell$'s and determine a band power within each $\ell$ bin.
In addition, to increase the signal to noise ratio
we use three bins $\ell=2$--$150$, $151$--$693$, and $\geq 694$ 
such that only the middle bin should have signal given the combination of beam size and sky area.

When estimating the band power and presenting the results
we must specify a model for the shape of the power spectrum within each $\ell$ bin.
Given the binned values, we model the power spectra as
\begin{mathletters}
\label{eq-shape}
\begin{equation}
	C_{\ell}^{YY'}= {\cal D}_{\ell\ell'} W_{\ell' b} c_b^{YY'},
\end{equation}
where
the subscript $b$ labels an $\ell$ bin,
both $C_{\ell}^{YY'}$ and $c_b^{YY'}$ are treated as column vectors,
${\cal D}_{\ell\ell'}$ is a square diagonal matrix with diagonal elements equal to
\begin{equation}
	\label{D}
	{\cal D}_{\ell\ell}=D_\ell,
\end{equation}
and $W_{\ell b}$ is a matrix defined as
\begin{equation}
	\label{W}
	W_{\ell b} =
		\left\{
		\begin{array}{ll}
			1, & {\rm when \;}\ell\in b, \\
			0, & {\rm when \;}\ell\notin b.
		\end{array}
		\right.
\end{equation}
Here $c_b^{YY'}$ is defined as the `band power',
and $D_\ell$ is called the `shape function'.
\end{mathletters}

The model for the shape is encoded in $D_\ell$.
We investigate the following four cases:
\begin{mathletters}
	\label{eq:shape}
\begin{eqnarray}
	{\textrm 1.} & D_\ell^{(1)}= & \frac{1}{\ell(\ell+1)}; \\
	{\textrm 2.} & D_\ell^{(2)}= & C_{\ell (\Lambda\rm CDM)}; \\
	{\textrm 3.} & D_\ell^{(3)}= & \frac{1}{2\ell+1};\\
	{\textrm 4.} & D_\ell^{(4)}= & 1.
\end{eqnarray}
Here the $C_{\ell (\Lambda\rm CDM)}$ is the power spectrum predicted by the concordance model
of the WMAP+ACBAR+BOOMERanG result in \cite{wmap-06cos}. %==========gap=====HERE
It is a flat $\Lambda$CDM cosmology
with $\Omega_bh^2=0.022$, $\Omega_mh^2=0.13$, $h=0.74$, $\tau=0.088$, $n_s=0.95$, and $\sigma_8=0.74$.
Consequently the power spectra for different models of the shape function
are related to the estimated band power as
\end{mathletters}
\begin{equation}
	\label{Dn}
	C_{\ell(n)}^{YY'}= {\cal D}_{\ell\ell'}^{(n)} W_{\ell' b} c_{b(n)}^{YY'},
\end{equation}
where $n=1,2,3,$ or $4$,
and $c_{b(n)}^{YY'}$ is the band power to be estimated.

\begin{figure}%[bt] 
\plotone{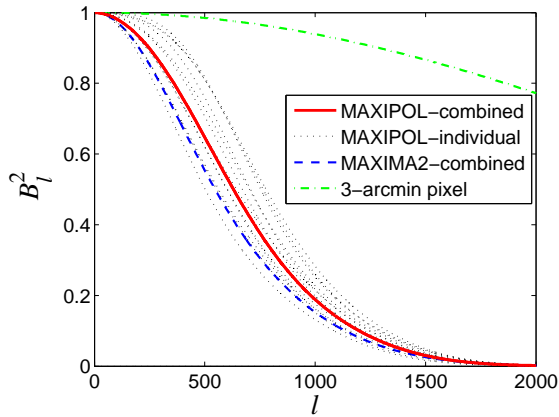}
  \caption
  {\small The $B_\ell^2$ function
  of each of the 12 MAXIPOL polarimeters (dotted), 
  of the noise-weighted combination (solid), 
  of the noise-weighted combination for MAXIMA2 (dashed) \citep{ma-abroe}, and 
  of the map pixelization (dot-dashed).}
  \label{fig-bl2}
\end{figure}

Prior to the estimation of the CMB power spectra,
we also need to determine the effects of the beam convolution and of the map pixelization,
so that these effects can be deconvolved during the estimation process.
We follow the recipes described in \citet{asym-beam}.
The two convolution effects are combined into a single transfer function $B_\ell$
in the multipole space:
\begin{equation}
	B_\ell^2 =  B_{\ell{\rm (beam)}}^2 B_{\ell{\rm(pxl)}}^2,
	\label{bl2}
\end{equation}
where $B_{\ell{\rm (beam)}}^2$ and $B_{\ell{\rm(pxl)}}^2$ are the effective $B_\ell^2$
of the beam and the pixelization, respectively.
The resulting $B_\ell^2$ for MAXIPOL are presented in Figure~\ref{fig-bl2}.
The individual $B_{\ell{\rm (beam)}}^2$ of the MAXIPOL polarimeters are somewhat smaller than those of MAXIMA
and thus produce a wider noise-weighted combination.
The figure also shows that with a pixel size of 3 arc-minutes
less than 2\% of signal power for $\ell \leq 700$ is attenuated by the pixelization.

%-------------------------------------------------FIG 5: the B_l^2 =========here

%%%%%%%%%%%%%%%%%%%%%%%%%%%%%%%%%%%%%%%%%%%%%%%%%%%%%%%%%%%%%%%%%%%%%%%%%%%%%%%%%%%%%%%%%%%%%%%%%%%%%%
%%%%%%%%%%%%%%%%%%%%%%%%%%%%%%%%%%%%%%%%%%%%%%%%%%%%%%%%%%%%%%%%%%%%%%%%%%%%%%%%%%%%%%%%%%%%%%%%%%%%%%
%%%%%%%%%%%%%%%%%%%%%%%%%%%%%%%%%%%%%%%%%%%%%%%%%%%%%%%%%%%%%%%%%%%%%%%%%%%%%%%%%%%%%%%%%%%%%%%%%%%%%%
\subsubsection{Bayesian Approach}
\label{ps-estimation-bayesian}

A commonly used Bayesian approach for power spectrum estimation is 
the Newton-Raphson algorithm \citep{bjk}.
We attempted to use this method to find the MAXIPOL power spectra,
but it failed in a variety of ways
that we believe were due to the low signal-to-noise ratio of our maps.
The region of parameter space we are exploring is close to the boundary where $C_\ell=0$, 
and the allowed solutions include negative power values.
This region has a non-smooth likelihood that makes the Newton-Raphson method unreliable.
We note that future B-mode experiments with a low signal to noise ratio in their B-mode detection are likely to face the same problem.
See also \citet{ma-abroe} for high signal-to-noise cases where the Newton-Raphson method is also prone to failure.
Thus we adopted a Markov Chain Monte-Carlo (MCMC) approach.

The MCMC method explores the likelihood space of the map, $P(C_\ell | \textrm{map})$.
It generates lists of samples from a parameter space 
whose distribution is asymptotically the same as the posterior distribution of the parameters.  
This approach has a number of advantages: 
it fully explores the parameter space, making no assumptions about the shape of the likelihood surface, 
and can be used in any signal-to-noise regime.  
The MCMC method is valid in cases
where the shape of the posteriors cannot be assumed to take simple forms, such as a Gaussian shape. 
It also has the important disadvantage of high computational cost.  

We used Metropolis-Hastings (MH) sampling, 
which is one of the simplest forms of MCMC. 
It has been widely used in cosmological parameter estimation \citep{wmap-06cos,B03-analysis,cosmomc}.
This method is also briefly mentioned in \citet{dasi-02}.
In our case the parameters are binned $C_\ell$ and the random variable is the map vector, $m$, made up of the $Q$ and $U$ pixel values.
We take the likelihood of the map vector to be Gaussian, 
with zero mean and inter-pixel covariance matrices  
$C=N+S(C_\ell)$, where $N$ is the noise covariance and $S(C_\ell)$ the signal correlation matrix
of the proposed $C_\ell$ \citep{tegmark}.
Thus the likelihood was calculated exactly using
$\log{L} = -\frac{1}{2}(m^T C^{-1} m + \mbox{Tr}~\log{C})$.
Our MCMC code was derived from the spectrum solver MADspec,
which is part of MADCAP \citep{madcap}.  The computation of the inverse matrix was the largest computational step and was done with a Cholesky decomposition.

To generate sufficient samples for estimating the parameters
we used two chains, each of approximately 50,000 samples.
The calculation required 24 hours on 128 processors on the Seaborg supercomputer, 
which belongs to the National Energy Research Scientific Computing Center
at Lawrence Berkeley National Laboratory.
We are in the process of optimizing the method 
and believe sufficient computational savings can be made to make the algorithm scalable to somewhat larger datasets.

Posterior likelihoods for binned $C_\ell$  can easily be calculated with the MH algorithm,
since the likelihood of a parameter value is proportional to its multiplicity in the chain. 
To perform these calculations we used the program GetDist, a part of the CosmoMC package \citep{cosmomc}, 
which also performs some convergence tests based on derived secondary chains.  
Additional convergence tests  based on power spectra of parameter values in the chain, proposed in \citet{dunkley}, were also performed.
In particular, the variance of the mean of the two chains was less than 10\% of the mean of their variances for the parameters of interest.

The choice of a good proposal density is critical to optimize the convergence rate of an MCMC chain.  
We followed the recommendations in \citet{cosmomc-notes}: 
We first re-parameterized the space by the eigenvectors 
of the parameter covariance matrix to mitigate the effect of highly correlated parameters.  
New proposed jumps were generated along orthonormal basis vectors of this new parameter set 
which were randomly rotated every $n_{bin}$ proposals. 
The length of the jump was a Gaussian random variable with mean zero and the variance of 
the appropriately rotated eigenvalue multiplied by a scaling factor of $2.4$.   
The covariance matrix was estimated with a short non-optimized MCMC.

%%%%%%%%%%%%%%%%%%%%%%%%%%%%%%%%%%%%%%%%%%%%%%%%%%%%%%%%%%%%%%%%%%%%%%%%%%%%%%%%%%%%%%%%%%%%%%%%%%%%%%
%%%%%%%%%%%%%%%%%%%%%%%%%%%%%%%%%%%%%%%%%%%%%%%%%%%%%%%%%%%%%%%%%%%%%%%%%%%%%%%%%%%%%%%%%%%%%%%%%%%%%%
%%%%%%%%%%%%%%%%%%%%%%%%%%%%%%%%%%%%%%%%%%%%%%%%%%%%%%%%%%%%%%%%%%%%%%%%%%%%%%%%%%%%%%%%%%%%%%%%%%%%%%
\subsubsection{Frequentist Approach}
\label{ps-estimation-frequentist}

\begin{figure*}%[bt] 
\plotone{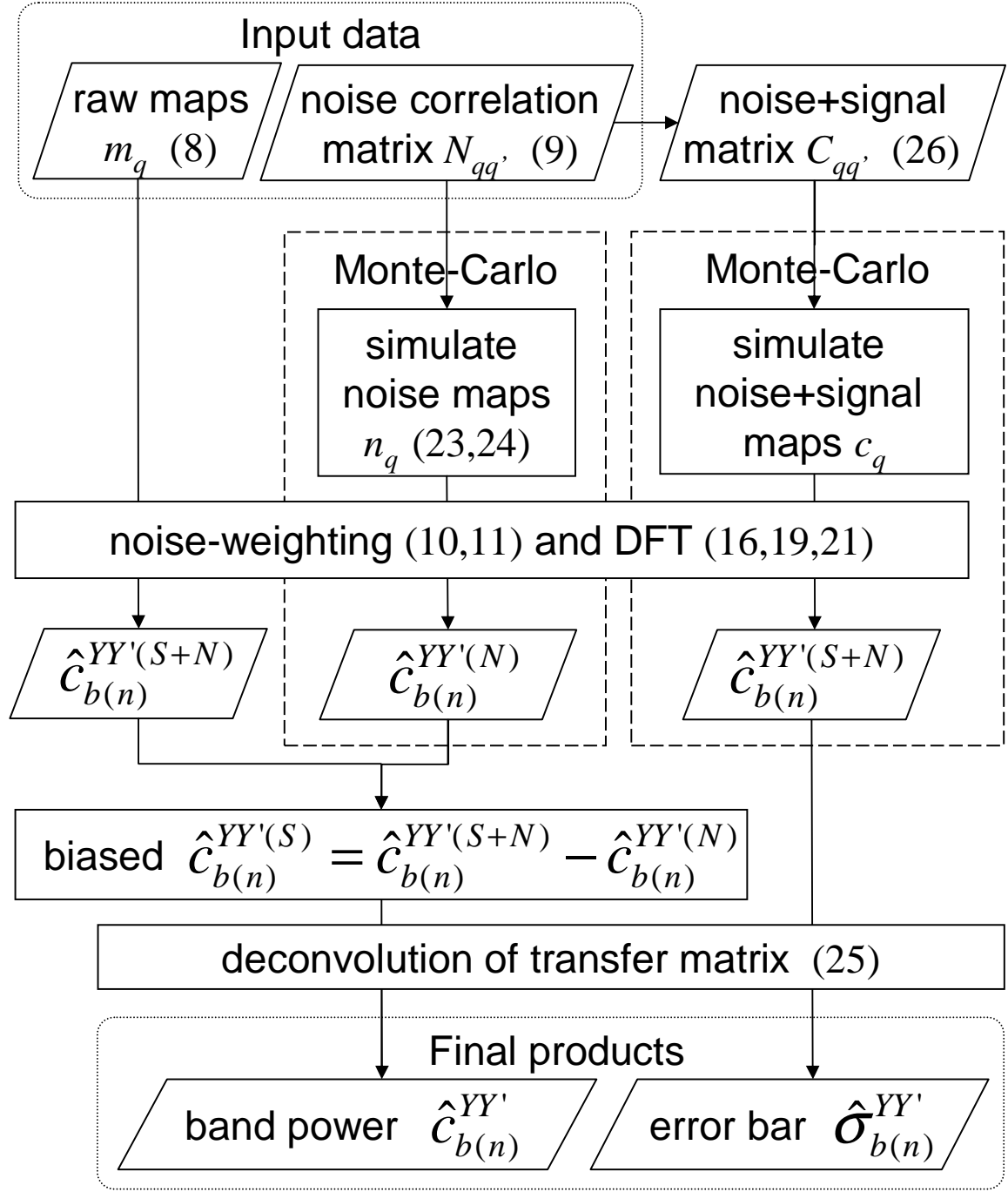}
  \caption
  {\small Flowchart of the frequentist approach for
	estimating the CMB power spectra and their associated errors.
	The numbers in brackets are the equation numbers associated with the operation. 
	The dashed boxes indicate the operations included in the Monte-Carlo simulation.
}
  \label{fig-flowchart}
\end{figure*}

We also used a frequentist approach to estimate $C_\ell^{YY'}$. 
Frequentist approach was used in the past in analyses of other data sets 
\citep{master,B03,wmap-06pol,xfaster}.  
The following sections describe the steps in our analysis and
Figure~\ref{fig-flowchart} shows a flowchart of the pipeline.

%-------------------------------------------------------FIG 6: FLOW CHART ====here

%--------------------------------------------------------flat-sky approximation
Because the sky patch of our observation is only about 4 degrees across
we approximate it as flat to speed up our computation.
We use a Discrete Fourier Transform (DFT) to approximate the multipole expansion.
With this approximation
equation~(\ref{QU-lm}) can be reduced and reorganized as
%\begin{mathletters}
\label{fft}
\begin{equation}
	\label{EB-k}
	\left(
		\begin{array}{c}
			a_{\bf k}^E \\
			a_{\bf k}^B
		\end{array}
	\right)
	=
	\left(
		\begin{array}{cc}
			-\cos 2\theta_{\bf k} & -\sin 2\theta_{\bf k}\\
			\sin 2\theta_{\bf k}  & -\cos 2\theta_{\bf k}
		\end{array}
	\right)
	\left(
		\begin{array}{c}
			\widetilde{Q}_{\bf k} \\
			\widetilde{U}_{\bf k}
		\end{array}
	\right),
\end{equation}
where a tilde denotes the Fourier coefficient of the corresponding quantity,
${\bf k}$ is a Fourier mode,
and $\theta_{\bf k}$ is the phase angle of ${\bf k}$.
The multipole number can thus be approximated as the wave number, $\ell \approx k \equiv |{\bf k}|$.
As a consequence, equation~(\ref{Cl-lm}) reduces to 
\begin{equation}
  \label{Cl-k}
  C_\ell^{YY'} \approx
  C_k^{YY'} =
	\frac{1}{{\cal A}}
	  \left\langle
		\left|
			a_{\bf k}^Y a_{\bf k}^{Y'*}
		\right|^2
	  \right\rangle_{|{\bf k}|=k},
\end{equation}
where ${\cal A}$ is the area of the map used in the DFT in steradians,
and the brackets denote an average over all the wave vectors with $|{\bf k}|=k$.

%--------------------------------------incorporating the shape function & band power
To increase the S/N ratio per $\ell$ bin in our results,
we combine the $\ell$'s into only three bins
and estimate their band powers $c_{b(n)}^{YY'}$, instead of the $C_\ell^{YY'}$,
with a specified shape function (see Eqs.~(\ref{eq:shape}) and (\ref{Dn})).
This requires a modification of equation (\ref{Cl-k}) as
\begin{equation}
  \label{Cl-k2}
  c_{b(n)}^{YY'} \approx
	\frac{1}{{\cal A}}
	  \left\langle
	  \frac{
	  		\left|
				a_{\bf k}^Y a_{\bf k}^{Y'*}
				\right|^2
		}{D_k^{(n)}}
	  \right\rangle_{|{\bf k}|\in b},
\end{equation}
where 
$D_k^{(n)}$ is the shape function $D_\ell^{(n)}$ 
convolved by the multipole transform of the sky-coverage window.
Note that prior to the DFT,
we apply the filtering of equations (\ref{weighting}) and (\ref{noise-weighting}) to the maps.

Due to the finite sky coverage, the flat-sky approximation, and the filtering of maps,
equation (\ref{Cl-k2}) is a biased estimator of the band powers.
This bias is corrected using the Monte-Carlo approach.

%===================================================================
%===================================================================
%===================================================================
\begin{center}
2.3.2.1 The Pseudo-Band Powers
\end{center}
%\subsubsubsection{\uppercase{T}he Pseudo-Band Powers}

We apply the DFT of Equations~(\ref{EB-k}) and (\ref{Cl-k2})
to each of the square regions selected from the original maps $m_p^X$ shown in Figure~\ref{fig-scan}
to obtain the estimated band powers $\hat{c}_{b(n)}^{YY'(S+N)}$,
where a hat denotes an estimator.
We refer to these as the `pseudo'-band powers,
because they are biased and contain noise.
These pseudo-band powers can be modelled as
\begin{mathletters}
\label{Cl-pseudo}
\begin{eqnarray}
	\left(
		\begin{array}{c}
			{c}^{EE(S+N)}_{b(n)} \\
			{c}^{BB(S+N)}_{b(n)}
		\end{array}
	\right)
	& = &
	H_{\{bb\}\{b'b'\}}^{(n)}
	\left(
		\begin{array}{c}
			c_{b'(n)}^{EE} \\
			c_{b'(n)}^{BB}
		\end{array}
	\right)  \nonumber \\
	& &
	+
	\left(
		\begin{array}{c}
			c_{b(n)}^{EE(N)} \\
			c_{b(n)}^{BB(N)}
		\end{array}
	\right),
	\label{Cl-EE-BB-pseudo} \\
	%-------------------------------------------------------------------------
	c_{b(n)}^{EB(S+N)}
	& = &
	\left(
		{_+}H^{(n)}_{bb'} - {_-}H^{(n)}_{bb'}
	\right)
	c_{b'(n)}^{EB} \nonumber \\
	& &
	+
	c_{b(n)}^{EB(N)},
	\label{Cl-EB-pseudo}
\end{eqnarray}
where 
%-------------------------------------------------------------------------
\begin{eqnarray}
	H_{\{bb\}\{b'b'\}}^{(n)}  &= &
	\left(
		\begin{array}{cc}
			{_+}F^{(n)}_{b\ell'} & {_-}F^{(n)}_{b\ell'} \\
			{_-}F^{(n)}_{b\ell'} & {_+}F^{(n)}_{b\ell'}
		\end{array}
	\right)\times \nonumber\\
	& &
	\left(
		\begin{array}{c}
			G_{\ell'\ell} B_{\ell}^2 W_{\ell b'} \\
			G_{\ell'\ell} B_{\ell}^2 W_{\ell b'}
		\end{array}
	\right), \\
  {_{\pm}}H^{(n)}_{bb'}  & =  &
  {_{\pm}}F^{(n)}_{b\ell'}G_{\ell'\ell} B_{\ell}^2 W_{\ell b'}.
\end{eqnarray}
Here 
the subscript $\{bb\}$ denotes the index of $b$ that runs twice,
and similarly for $\{\ell\ell\}$.
The $c_{b(n)}^{YY'}$ are the underlying CMB signals of the field,
and the $c_{b(n)}^{YY'(N)}$ are the noise components of the pseudo-band powers.
The $B_{\ell}$, $G_{\ell'\ell}$, and ${_{\pm}}F^{(n)}_{b\ell'}$ will be explained as follows.
\end{mathletters}

%-------------------------------------------------------------------------------------------B_l
We call $B_\ell$ the `beam transfer function', which has been discussed earlier.
It accounts for the convolution effects
from the beam pattern of each of the polarimeters
and
from the pixelization during the map-making process.

%-------------------------------------------------------------------------------------------G_ll
We call the $G_{\ell\ell'}$ in equations (\ref{Cl-pseudo}) the `time-domain transfer matrix'.
It is induced by the time-domain processing,
including the effects from the demodulation and the deglitching of the data.
Here we use the same $G_{\ell\ell'}$ for all the $EE$, $EB$, and $BB$ modes
because a signal-only Monte-Carlo simulation of 1,000 realizations indicate that
it is the same to an accuracy of $2\%$.
This is essentially because 
the form of $G_{\ell\ell'}$ is dominated by the band-pass filter in the demodulation process
and thus behaves simply as a convolution effect of the sky signal.

%-------------------------------------------------------------------------------------------F_ll
We call ${_{\pm}}F^{(n)}_{b\ell'}$ in equations (\ref{Cl-pseudo}) the `DFT transfer matrix'.
It accounts for the biasing effect from the DFT approach
(Eqs.~(\ref{weighting}), (\ref{noise-weighting}), (\ref{EB-k}), and (\ref{Cl-k2})).
The ${_+}F^{(n)}_{b\ell'}$ and ${_-}F^{(n)}_{b\ell'}$ are dominated by 
the self-coupling and the geometric mixing of E and B polarization
respectively \citep{chon}.

With the  formalism (\ref{Cl-pseudo}) established, 
our task becomes to obtain an unbiased estimator of $c_{b(n)}^{YY'}$.
This requires a inversion of equations (\ref{Cl-EE-BB-pseudo}) and (\ref{Cl-EB-pseudo}).
As will be shown,
the forms of $G_{\ell\ell'}$ and ${_{\pm}}F^{(n)}_{b\ell'}$ do not need to be estimated individually.
Instead, we estimate the overall transfer matrices $H_{\{bb\}\{b'b'\}}^{(n)}$ 
and $\left({_+}H^{(n)}_{bb'} - {_-}H^{(n)}_{bb'}\right)$
given a specified shape function $D_{\ell}^{(n)}$ and the measured $B_\ell$,
and then compute their inverses.

%===================================================================
%===================================================================
%===================================================================
\begin{center}
2.3.2.2 Estimation of the Noise Component
\end{center}
%\subsubsubsection{\uppercase{E}stimation of the noise component}

Following the frequentist approach,
we estimate the noise component $c^{YY'(N)}_{b(n)}$ in equations~(\ref{Cl-pseudo})
by using the previously estimated pixel-pixel noise correlation matrix
$N_{qq'}$ to carry out a Monte-Carlo simulation for the noise-only maps.
A Cholesky decomposition gives
\begin{equation}
	N_{qq'}=LL^{\rm T},
	\label{chol}
\end{equation}
where $L$ is a lower triangular matrix.
Then one realization of the simulated noise map is obtained by taking
\begin{equation}
	n_q
	=
	\left(
		\begin{array}{c}
			n_p^{Q} \\
			n_p^{U}
		\end{array}
	\right)
	=Lg,
	\label{mc}
\end{equation}
where $g$ is a vector of Gaussian random numbers with mean zero and variance one.
Finally, applying the DFT approach 
(Eqs.~(\ref{weighting}), (\ref{noise-weighting}), (\ref{EB-k}), and (\ref{Cl-k2})) 
to all these noise maps yields an estimated $\hat{c}_{b(n)}^{YY'(N)}$.
We use 10,000 realizations for the Monte-Carlo to obtain our results.

%===================================================================
%===================================================================
%===================================================================
\begin{center}
2.3.2.3 Unbiased Estimator
\end{center}
%\subsubsubsection{\uppercase{U}nbiased estimator}

We now construct an unbiased estimator for $c_{b(n)}^{YY'}$.
Taking the inverse operation of equation (\ref{Cl-EE-BB-pseudo}) gives
\begin{mathletters}
\begin{eqnarray}
	\left(
		\begin{array}{c}
			\hat{c}_{b(n)}^{EE} \\
			\hat{c}_{b(n)}^{BB}
		\end{array}
	\right)
	= 
	\left({\hat{H}^{(n)}_{\{bb\}\{b'b'\}}}\right)^{-1}
	\times  \nonumber \\
	\left(
		\begin{array}{c}
			\hat{c}_{b'(n)}^{EE(S+N)} - \hat{c}_{b'(n)}^{EE(N)} \\
			\hat{c}_{b'(n)}^{BB(S+N)} - \hat{c}_{b'(n)}^{BB(N)}
		\end{array}
	\right),
	\label{Cb-EE-BB-est}
\end{eqnarray}
and similarly
\begin{eqnarray}
	\hat{c}_{b(n)}^{EB}
	= &
	\left(
		{_+}\hat{H}_{bb'}^{(n)} - {_-}\hat{H}_{bb'}^{(n)}
	\right)^{-1} 
	\times  \nonumber \\
	&
	\left(
		\hat{c}_{b'(n)}^{EB(S+N)} - \hat{c}_{b'(n)}^{EB(N)}
	\right).
	\label{Cb-EB-est}
\end{eqnarray}
The inversion operation for $\hat{H}_{\{bb\}\{b'b'\}}^{(n)}$ and 
$\left({_+}\hat{H}_{bb'}^{(n)} - {_-}\hat{H}_{bb'}^{(n)}\right)$ here
is feasible only if the underlying matrices are square, 
i.e., only if the numbers of $b$'s and $b'$'s are the same.
We thus use the same binning strategy for $b$ and $b'$
with only three wide bands in obtaining our results.
\label{Cb-est}
\end{mathletters}

%------------------------------------------Monte-Carlo to get H
To estimate $H_{\{bb\}\{b'b'\}}^{(n)}$, we employ the following end-to-end Monte-Carlo simulation.
We inject a unit power into $c_{b(n)}^{YY'}$ for one $\ell$ bin $b$ at a time.
After multiplying the resulting $C_{\ell(n)}^{YY'}$ obtained from equation~(\ref{Dn})
with $B_{\ell {\rm(beam)}}^2$,
we use equation~(\ref{QU-lm}) to obtain the signal-only high-resolution maps of $Q$ and $U$,
which are then scanned and processed to produce mock TOPD.
Maps computed from equation~(\ref{mp}) using the noise matrices measured from the real data
are processed through equations~(\ref{weighting}), (\ref{noise-weighting}), (\ref{EB-k}), and (\ref{Cl-k2})
to yield the resulting band powers.
These band powers give one column of the transfer matrix $H_{\{bb\}\{b'b'\}}^{(n)}$
that corresponds to the chosen $\ell$ bin for input.
A Monte-Carlo simulation with 1,000 realizations is used to
obtain each of the six columns in $H_{\{bb\}\{bb'\}}^{(n)}$.

Finally an inversion of $H_{\{bb\}\{bb'\}}^{(n)}$
and the previously estimated noise components 
are used in equations (\ref{Cb-est})
to yield unbiased estimates of the band powers $c_{b(n)}^{YY'}$
and thus the power spectra $C_{\ell(n)}^{YY'}$.

%===================================================================
%===================================================================
%===================================================================
\begin{center}
2.3.2.4 Estimation of Error Bars
\end{center}
%\subsubsubsection{\uppercase{E}stimation of error bars}

To estimate the error bars of the power spectra,
we again employ a Monte-Carlo simulation.
First, we simulate maps that contain both signal and noise
using equations (\ref{chol}) and (\ref{mc}),
but with the $N_{qq'}$ replaced by 
\begin{equation}
	\label{Cqq}
	C_{qq'}=S_{qq'} + N_{qq'}.
\end{equation}
Here $S_{qq'}$ is the signal-signal correlation matrix based on the $\hat{c}_{b(n)}^{YY'}$.
The use of the DFT approach and equations (\ref{Cb-est}) yields the band powers.
We compute 10,000 realizations of such band powers, 
obtain the probability distribution for the power value within each $\ell$ bin $b$, 
and calculate the 68\% confidence intervals.

%%%%%%%%%%%%%%%%%%%%%%%%%%%%%%%%%%%%%%%%%%%%%%%%%%%%%%%%%%%%%%%%%%%%%%%%%%%%%%%%%%%%%%%%%%%%%%%%%%%%%%
%%%%%%%%%%%%%%%%%%%%%%%%%%%%%%%%%%%%%%%%%%%%%%%%%%%%%%%%%%%%%%%%%%%%%%%%%%%%%%%%%%%%%%%%%%%%%%%%%%%%%%
%%%%%%%%%%%%%%%%%%%%%%%%%%%%%%%%%%%%%%%%%%%%%%%%%%%%%%%%%%%%%%%%%%%%%%%%%%%%%%%%%%%%%%%%%%%%%%%%%%%%%%
\section{Results}
\label{results}

%%%%%%%%%%%%%%%%%%%%%%%%%%%%%%%%%%%%%%%%%%%%%%%%%%%
\subsection{Maps and power spectra}
\label{results-maps-ps}

\begin{figure*}
  \centering
	\includegraphics[width=140mm]{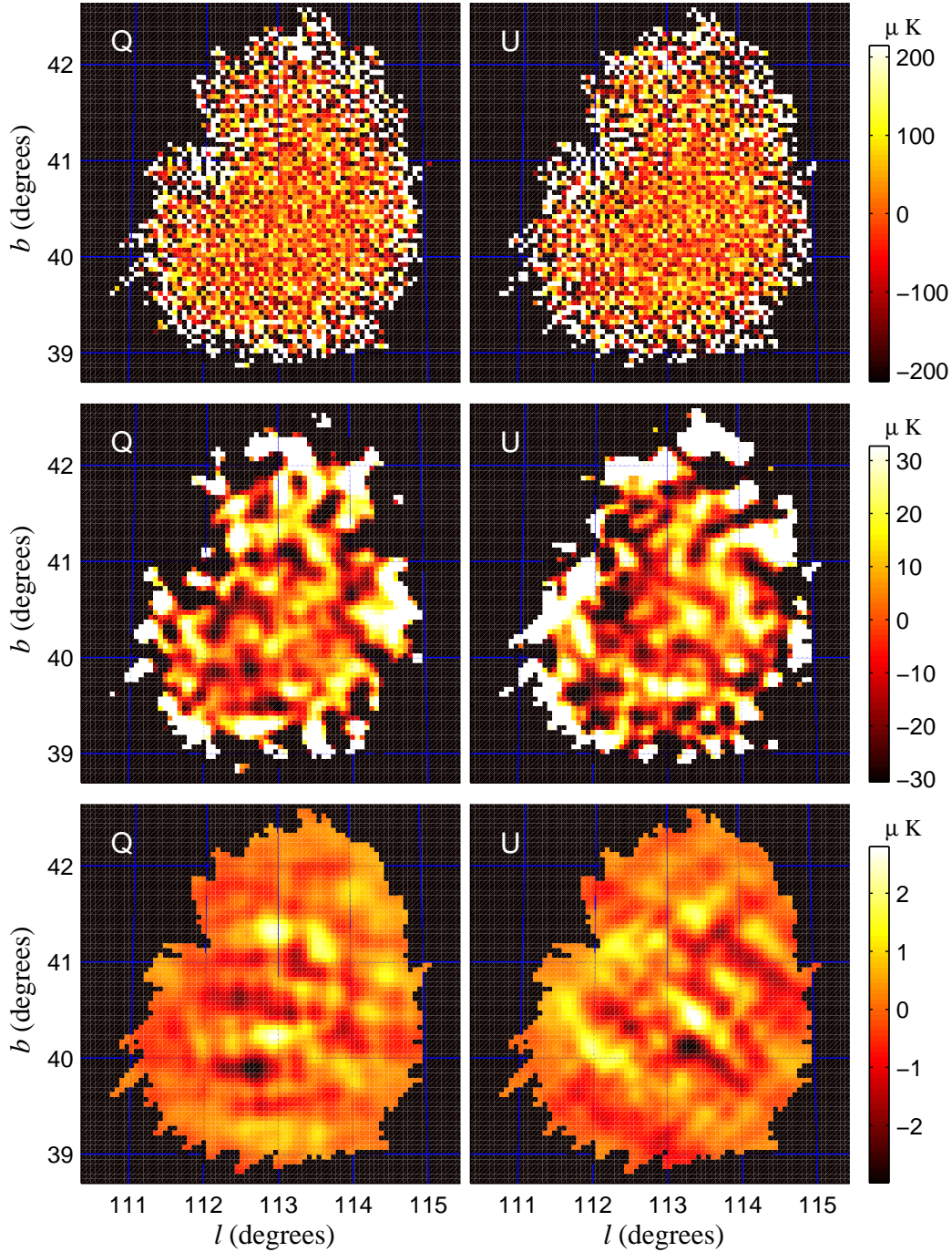}
  \caption{\small The maps of MAXIPOL polarization data (top),
  the maps convolved with a 10-arcmin Gaussian beam (middle), and
  the Wiener-filtered maps (bottom). 
}
  \label{fig-maps}
\end{figure*}

%We processed the TOD to obtain the $Q$ and $U$ maps and the associated noise correlation matrix,
%using the demodulation technique and the maximum-likelihood method 
%described in Sections~\ref{lock-in} and \ref{map-making} respectively.
Figure~\ref{fig-maps} shows
the maps $m_p^X$ for the $Q$ and $U$ Stokes parameters,
the maps convolved with a 10-arcmin Gaussian beam, and 
the Wiener-filtered maps (see Eqs.~(\ref{weighting}) and (\ref{wiener-filtering})). 
Note that the three sets have different color ranges. 
Some uncleaned systematic errors in the TOD, such as long-baseline gain drift,  can manifest as non-Gaussian map components, such as scan-striping or gradients.
It is clear that 
none of the maps show such visible evidence for systematic errors.
We will conduct more tests for systematic errors in Section~\ref{results-check}.

%--------------------------------------FIG 7: Q,U maps: raw, conv., Wiener filtered===here

%-------------------------------------------- ps: frequentist + Bayesian + shape + calibration
For determining the amplitude of the power spectra we used a square region of 2.3 degrees across 
centered at $l=113^\circ$, $b=40.25^\circ$ (see Fig.~\ref{fig-scan}) and three $\ell$ bins:
$\ell=2$--$150$, $151$--$693$, and $\geq 694$.  
Therefore there are in total nine $\ell$ bins under consideration, 
three for each of the $EE$, $EB$, and $BB$ modes.  
We report results only for the central bin, unless otherwise stated. 
This is the only bin likely to contain signal, given the beam size and area of the maps. 
All the Bayesian results are reported after marginalization of the joint posterior likelihood
over un-interesting bins. 
Because there is no tractable Bayesian method to account for the loss of power 
due to the band-pass filtering during demodulation of the time streams,
we rescaled results of $C_\ell$ for the central $\ell$ bin by a factor of 1.06, 
calculated from an appropriate frequentist approach.
Bayesian results are quoted as the mode of the likelihood function 
with 68\% intervals of maximum likelihood. 
Frequentist results are quoted as the median of the probability distribution function
with 68\% intervals about the median.
We used a shape function $D_\ell^{(1)}=1/[\ell(\ell+1)]$ unless otherwise noted. 

Table~\ref{tab-shape} gives the amplitude of the polarization power spectra
using both analysis methods and for different shape functions $D_\ell^{(n)}$. 
For ease of direct comparison  we present all results in $\ell(\ell+1) C_\ell / (2\pi)$ 
at the bin center $\ell=422$.  
The table also gives the results after marginalizing over a calibration uncertainty 
that is assumed Gaussian with $\sigma_{\rm cal} = 13\%$.
The Bayesian and frequentist approaches give consistent results.
Results between different shape functions are also consistent within statistical uncertainties.

We note that our Bayesian results are more dependent on shape function than the frequentist results, with a variation approximately one sigma. Such results have been observed by other others, such as  in \citet{B03}, where the variation is as large as four sigma.  In our frequentist results the effect is less apparent, mainly because of the Monte-Carlo bias correction in our pipeline.  The origin of this  subtle effect warrants further investigation.

%--------------------------------------------Table 1: ps for B & F ===========here
\begin{table}
	\centering
	\caption{Amplitude of power spectra} %=======================here
	\begin{tabular}{c | ccc}
		\hline\hline
		Shape $D_\ell^{(n)}$	\T\B	& $EE$	& $EB$	& $BB$	\\
		\hline\hline
		${1}/{\ell(\ell + 1)}$\T\B  & $\bf 55^{+51}_{-45}$& $\bf 18^{+27}_{-34}$ & $\bf -31^{+31}_{-19}$ \\
		$\Lambda$CDM    			\T\B  & $109^{+130}_{-101}$  & $23^{+37}_{-37}$ & $-48^{+40}_{-25}$  \\
		${1}/{(2\ell + 1)}$   \T\B  & $83^{+68}_{-50}$    & $24^{+40}_{-38}$ & $-51^{+37}_{-20}$  \\
		$1$     							\T\B  & $117^{+61}_{-77}$   & $32^{+52}_{-40}$ & $-41^{+34}_{-33}$  \\
		\hline\hline
		\multicolumn{4}{c}{Bayesian approach (inc.\ $\sigma_{\rm cal}$)\T\B}\\
		\hline
		${1}/{\ell(\ell + 1)}$\T\B  & $53^{+57}_{-45}$    & $14^{+33}_{-31}$ & $-30^{+34}_{-21}$  \\
		$\Lambda$CDM    			\T\B  & $113^{+136}_{-109}$ & $27^{+36}_{-42}$ & $-41^{+39}_{-34}$  \\
		${1}/{(2\ell + 1)}$   \T\B  & $88^{+70}_{-58}$    & $29^{+39}_{-45}$ & $-41^{+34}_{-31}$  \\
		$1$  									\T\B  & $108^{+80}_{-72}$   & $34^{+53}_{-46}$ & $-47^{+46}_{-29}$  \\
		\hline\hline
		\multicolumn{4}{c}{Frequentist approach\T\B}\\
		\hline
		${1}/{\ell(\ell + 1)}$\T\B 	& $62^{+52}_{-45}$	& $3^{+33}_{-32}$		& $26^{+45}_{-50}$  \\
		$\Lambda$CDM 					\T\B	& $68^{+46}_{-45}$	& $5^{+34}_{-31}$		& $21^{+47}_{-38}$  \\
		${1}/{(2\ell + 1)}$ 	\T\B	& $73^{+69}_{-43}$	& $23^{+31}_{-48}$	& $38^{+42}_{-50}$  \\
		$1$ 									\T\B	& $72^{+75}_{-46}$	& $8^{+46}_{-42}$		& $21^{+66}_{-58}$  \\
		\hline\hline
	\end{tabular}
%	\caption{\small %===============here
\tablecomments{
	Amplitude of power spectra $\ell(\ell+1)C^{YY'}_\ell / (2\pi)$ in $\mu\mbox{K}^2$ 
	for a wide band $\ell=151$--$693$ assuming different shape functions $D_\ell^{(n)}$.
	Errors are 68\% confidence intervals.
	The middle block with `(inc.\ $\sigma_{\rm cal}$)' gives results 
	including calibration uncertainty $\sigma_{\rm cal} = 13\%$. %(see Sec.~\ref{results-onetail}).
	The first row (bold) is the result 
	for which we show the likelihoods in 
	Figures~\ref{fig-mcmc1} and~\ref{fig-mcmc2}, 
  and is the result that is shown in Figure~\ref{fig-ps}.
	}
	\label{tab-shape}
\end{table}

%---------------------------------------------- FIG 8: likelihood: 1D ===========here
\begin{figure}
	\plotone{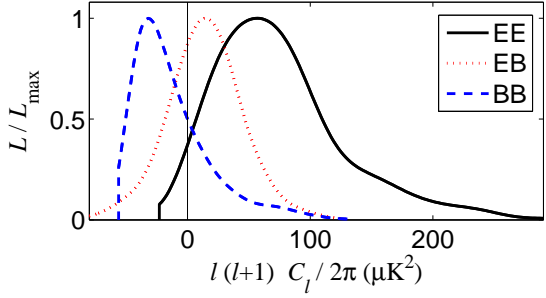}
	\caption{\small
	One-dimensional marginalized posterior likelihoods of $\ell(\ell+1) C^{YY'}_\ell / (2\pi)$.
	The sharp cut-offs at the negative end reflect the high	dimensionality of the marginalized space,
	and do not bias the analysis (see text).
	}
	\label{fig-mcmc1}
\end{figure}
 
%----------------------------------------------------- FIG 9: likelihood: 2D =====here
\begin{figure}
	\plotone{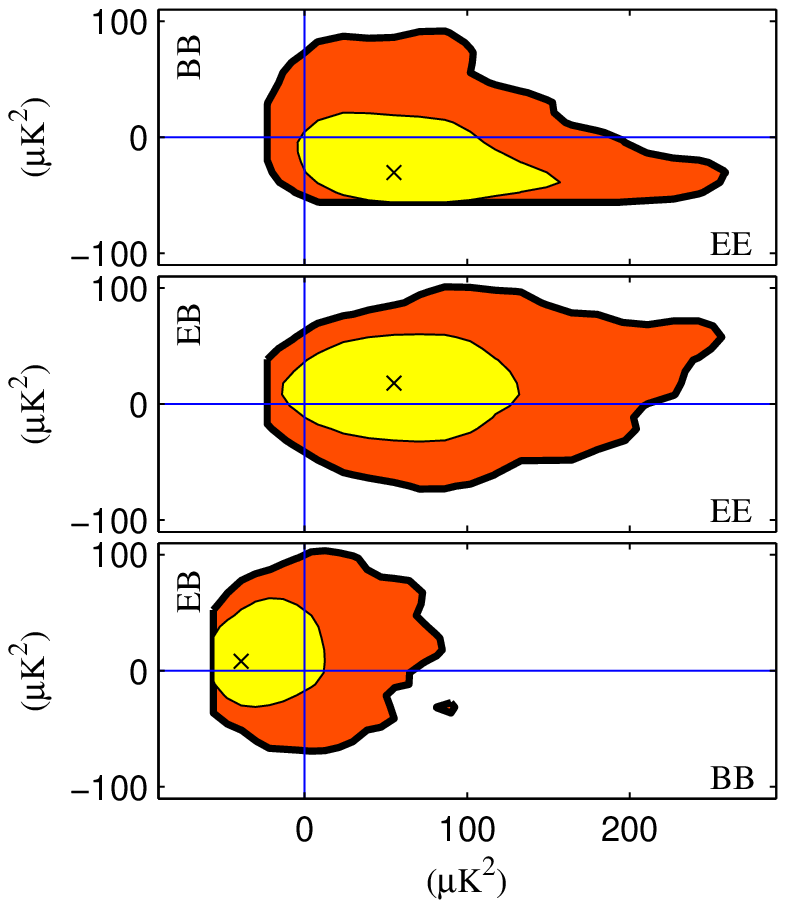}
	\caption{\small Two-dimensional marginalized posterior likelihoods of $\ell(\ell+1) C^{YY'}_\ell / (2\pi)$.
	The inner (thinner) and outer (thicker) contours indicate the confidence regions of $68\%$ and $95\%$ respectively.
	The crosses mark the locations of maximum likelihood.
	}
	\label{fig-mcmc2}
\end{figure}

%%%%%%%%%%%%%%%%%%%%%%%%%%%%%%%%%%%%%%%%%%%%%%%%%%%    ps-bayesian
%================================== EE
Figure~\ref{fig-mcmc1} gives the Bayesian posterior likelihoods of the $EE$, $EB$, and $BB$ modes.
Two of the likelihoods ($EE$ and $BB$) exhibit a sharp cut-off 
at the negative end of the parameter axis.
The cut-off arises when we marginalize the smooth multi-dimensional likelihood 
over the eight other parameters (bins).
Our likelihood computation method fails (and consequently we have  
zero samples) when any one eigenvalue of the total covariance matrix  
becomes negative.  Since multiple bins contribute to each eigenmode  
the cut-offs in the different parameters are coherent, and a sharp  
edge is formed on marginalization.
Because of such correlations between the cut-offs of different parameters,
integrating over $n$ parameters can lead to a cut-off that scales as $C_\ell^{n-1}$.  
Therefore the sharpness reflects the high dimensionality of the marginalized space.
It does not bias our analysis.

The posterior likelihood for $EE$ is skewed positive. 
By integrating the likelihood with a uniform prior over both positive and negative values 
we found a 96\% probability that $EE$ power is positive. 
This probability value is unchanged after inclusion of a 13\% calibration uncertainty. 

%================================== EB, BB
The posterior likelihoods for $EB$ and $BB$ are consistent with no signal. 
The $95\%$ confidence intervals for the two modes are
$-53~\mu\mbox{K}^2 \leq \ell(\ell+1)C^{EB}_\ell/(2\pi) \leq 81~\mu\mbox{K}^2$ and 
$-55~\mu\mbox{K}^2 \leq \ell(\ell+1)C^{BB}_\ell/(2\pi) \leq 57~\mu\mbox{K}^2$, respectively.  
To obtain an upper limit for the $BB$-mode we removed the negative region of its likelihood 
(see Fig.~\ref{fig-mcmc1}) and renormalized the rest.  
We found that at the $95\%$ confidence level
$[\ell(\ell+1)C^{BB}_\ell/(2\pi)]^{1/2}\leq 10.6$ and $9.5~\mu\mbox{K}$ 
with and without the inclusion of calibration uncertainty respectively.

Figure~\ref{fig-mcmc2} shows Bayesian two-dimensional joint posteriors. 
Joint distributions that include the $EE$ mode are skewed positive in $EE$.
There are sharp cut-offs of the joint likelihood surface for the same reason
that they occur in the one-dimensional likelihoods.
This results in straight edges for some of the 95\% confidence contours.

\begin{figure*}
  \centering
	\includegraphics[width=130mm]{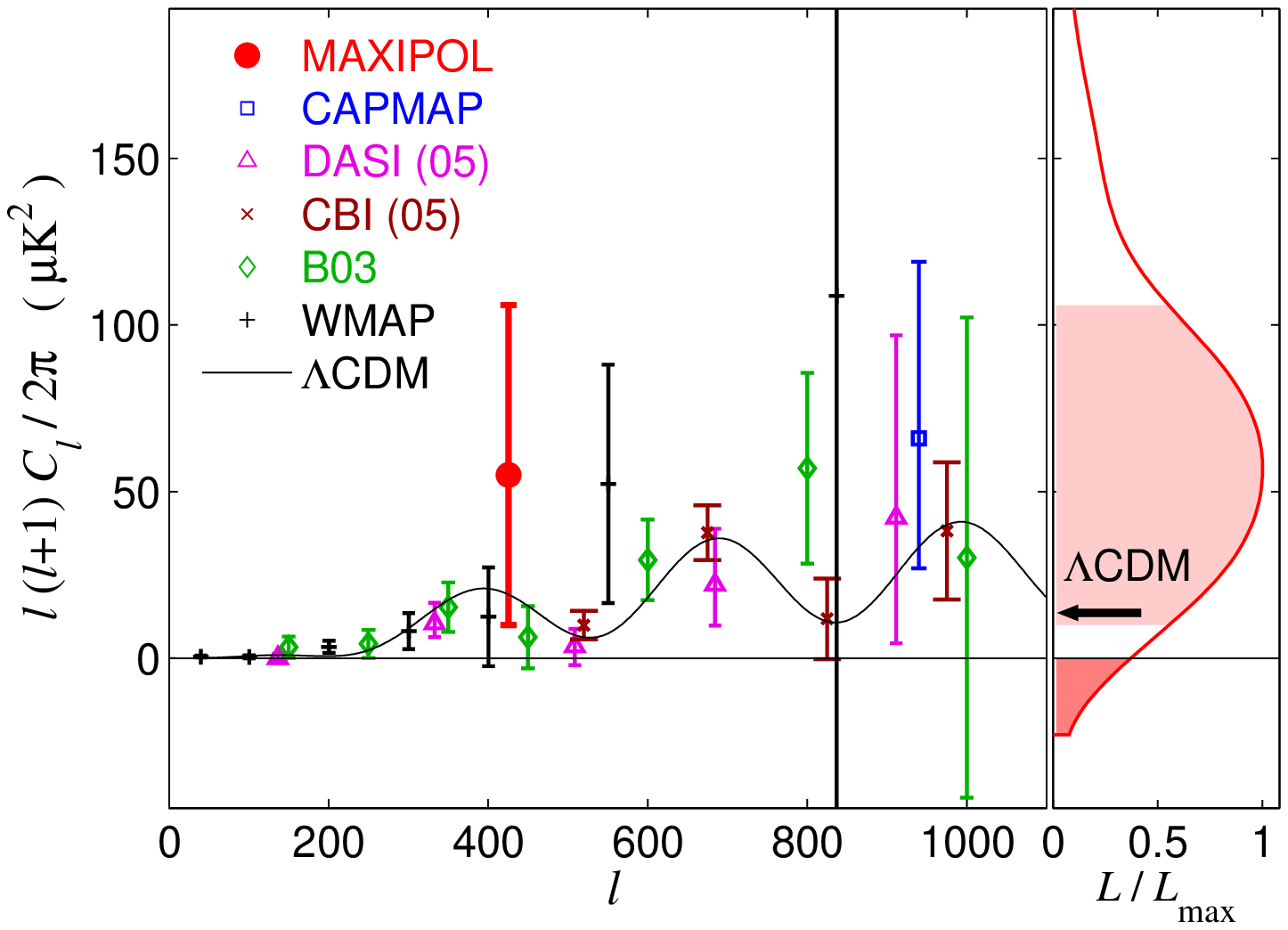}
	\caption{\small The $\ell(\ell+1)C^{EE}_\ell/(2\pi)$ at $\ell \leq 1000$ from all experiments 
		which reported CMB polarization.
		Our result is taken from the first row of Table~\ref{tab-shape} (numbers in bold type).
		The other results are from CAPMAP \citep{capmap-04}, DASI \citep{dasi-05}, B03 \citep{B03},
		CBI \citep{cbi-05}, and WMAP \citep{wmap-06pol}.
		The right panel is the posterior likelihood of our result.
		The solid curve in the left panel is a flat $\Lambda$CDM model from the WMAP+ACBAR+BOOMERanG
		result in \citet{wmap-06cos} (see text for details).
		It has a mean value of $\ell(\ell+1)C^{EE}_\ell/(2\pi)=14~\mu\mbox{K}^2$ for our bin at 
		$\ell=151$--$693$ (shown as an arrow in the right panel).
		The shaded regions in the right panel indicate the $68\%$ confidence region
		of the posterior likelihood and the region with negative $EE$ power, 
		which is 4\% of the total area below the likelihood curve.
	}
	\label{fig-ps}
\end{figure*}

%===================================== EE comparison with other CMB exp's.
We compare the amplitude of the $EE$ power spectrum
with results from other experiments in Figure~\ref{fig-ps}.  
The right panel in the figure is the same posterior likelihood as shown in Figure~\ref{fig-mcmc1}.
Our result is consistent with the prediction of the concordance model, 
which has a mean value of $\ell(\ell+1)C^{EE}_\ell/(2\pi)=14~\mu\mbox{K}^2$ for our $\ell$ bin.
This value falls at the 65\% confidence boundary of our likelihood function around the mode.
The lighter shaded region in the right panel of the figure indicates 
the $68\%$ confidence region of the posterior likelihood. 
The darker shaded region shows the area under the likelihood 
where $EE$ is negative, containing 4\% of the total area under the curve.

%---------------------------------------FIG 10: ps vs. other CMB exp. + likelihood-1D ====here

%%%%%%%%%%%%%%%%%%%%%%%%%%%%%%%%%%%%%%%%%%%%%%%%%%%
\subsection{Significance of the measured power}
\label{results-onetail}

%---------------------------------------------------------------EB=BB=0: power spectra
According to standard cosmological models 
$C_\ell^{EB}$ and $C_\ell^{BB}$ are predicted to be about one order of magnitude
smaller than the $C_\ell^{EE}$ and thus undetected by MAXIPOL.
We computed the Bayesian posterior likelihoods for $\ell(\ell+1)C^{EE}_\ell/(2\pi)$
with a prior $C_\ell^{EB}=C_\ell^{BB}=0$.   
The results with and without calibration uncertainty are summarized in Table~\ref{tab-eeonly}.
A comparison with Table~\ref{tab-shape} shows that including the priors
gives somewhat smaller modes and error bars.

%--------------------------------------------Table 2: ps for B & F (EB=BB=0) ===========here
\begin{table}
	\centering
	\caption{EE power spectrum with $C_\ell^{EB}=C_\ell^{BB}=0$} %==================here
	\begin{tabular}{c | ccc}
		\hline\hline
		Shape	$D_\ell^{(n)}$\T\B	& Mode	&	68\%					&	95\%\\
		\hline\hline
		\multicolumn{4}{c}{Bayesian $C_\ell^{EB}=C_\ell^{BB}=0$\T\B}\\
		\hline
		$1/[\ell(\ell + 1)]$\T\B 	& $12$	&	$^{+40}_{-21}$	& $^{+83}_{-38}$\\
		$1/(2\ell + 1)$ 		\T\B	& $41$	&	$^{+59}_{-38}$	& $^{+130}_{-71}$\\
		\hline\hline
		\multicolumn{4}{c}{Bayesian $C_\ell^{EB}=C_\ell^{BB}=0$ (inc.\ $\sigma_{\rm cal}$)\T\B}\\
		\hline
		$1/[\ell(\ell + 1)]$\T\B 	& $12$	&	$^{+41}_{-22}$	& $^{+94}_{-38}$\\
		$1/(2\ell + 1)$ 		\T\B	& $49$	&	$^{+56}_{-48}$	& $^{+144}_{-80}$\\
		\hline\hline
	\end{tabular}
%	\caption{\small %===============here
\tablecomments{
		Amplitude of power spectrum $\ell(\ell+1) C^{EE}_\ell / (2\pi)$ in $\mu\mbox{K}^2$, 
		and $68\%$ and $95\%$ confidence intervals assuming different shape functions
		$D_\ell^{(n)}$ with a prior $C_\ell^{EB}=C_\ell^{BB}=0$.
		}
	\label{tab-eeonly}
\end{table}

%-----------------------------------------------------------------------EB=BB=0: EE>0
Table~\ref{tab-eezero} summarizes the confidence level 
at which the hypothesis $C_\ell^{EE}=0$ is rejected for different shape functions,
and with and without a prior $C_\ell^{EB}=C_\ell^{BB}=0.$
The procedure for the calculation is identical to 
the one already discussed in Section~\ref{results-maps-ps}.  
We assume a uniform prior for all values of $EE$ and 
integrate the area below the appropriate likelihood function on the positive side.
Because the posterior likelihoods are all skewed positive 
most of the confidence levels for a positive $C_\ell^{EE}$ are above 90\%. 
These numbers do not depend on the magnitude of the calibration uncertainty
because the calibration uncertainty is a multiplicative factor,
which does not change the fraction of area under the likelihood for values $C_\ell^{EE}>0$.

%--------------------------------------------Table 3: Prob(EE>0) ===========here
\begin{table}
	\centering
	\caption{Probability for $C_\ell^{EE} >0$} %================here
	\begin{tabular}{c | c | c}
		\hline\hline
		Shape	$D_\ell^{(n)}$\T\B	& $\quad$ No Prior	$\quad$ & $C_\ell^{EB}=C_\ell^{BB}=0$	\\
		\hline
		$1/[\ell(\ell + 1)]$\T 	& 96\%	& 83\%  \\
		$\Lambda$CDM 					& 94\%	& -  \\
		$1/(2\ell + 1)$ 			& 98\%	& 92\%  \\
		$1$ 									& 98\%	& -  \\
		\hline\hline
	\end{tabular}
%	\caption{\small  %===============here
\tablecomments{
	Proportion of likelihood with positive $C_\ell^{EE}$ for
	different shape functions $D_\ell^{(n)}$ with no prior,
	and with a prior $C_\ell^{EB}=C_\ell^{BB}=0$ 
	(not all shape functions were considered in this case).
		}
	\label{tab-eezero}
\end{table}

%%%%%%%%%%%%%%%%%%%%%%%%%%%%%%%%%%%%%%%%%%%%%%%%%%%
\subsection{Systematic error tests}
\label{results-check}

%======================================================================difference maps
\subsubsection{Difference maps}
\label{sec:diff}
%----------------------------------------------------------------------sum/diff 1
We divided the TOD into two halves in the time domain and processed them separately
to yield the CMB maps $m_q^{\rm (h1)}$ and $m_q^{\rm (h2)}$.
Separate noise correlation matrices were computed.
The difference maps
\begin{equation}
	\label{mq-diff}
	m_q^{\rm (dif)} = m_q^{\rm (h1)} - m_q^{\rm (h2)},
\end{equation}
were constructed
and the resulting noise matrices computed.
We then estimated the polarization power spectra based on these $Q$ and $U$ maps,
using both Bayesian and frequentist approaches.
The rows labeled `time' in Table~\ref{tab-diff} show the results
with and without calibration uncertainty.
All these results are consistent with zero.  

%----------------------------------------------------------------------sum/diff 2
In a similar manner, 
we combined half of the 12 polarimeters to make one set of maps, 
and the other half for another set, 
and computed difference maps and the associated noise matrices.
The row labeled `polar' in Table~\ref{tab-diff} shows the results.

Within statistical uncertainties neither the time-domain differencing nor 
the polarimeter differencing test gives evidence for systematic errors.
We note that the sizes of the 68\% confidence intervals in Table~\ref{tab-diff}
are on average larger than those of the $EB$ and $BB$ modes in Table~\ref{tab-shape}
because the differencing process inevitably increases the noise level per pixel.

%--------------------------------------------Table 4: ps for diff maps ===========here
\begin{table}
	\centering
	\caption{power spectra of difference maps} %======================here
	\begin{tabular}{cc | ccc}
		\hline\hline
		Test\T\B	& Appr.	& $EE$	& $EB$	& $BB$	\\
		\hline
		time	\T\B	& B	& $-19^{+54}_{-29}$	& $-55^{+33}_{-38}$	& $8^{+45}_{-48}$	\\
		time	\T\B	& B ($\sigma_{\rm cal}$)	& $-10^{+48}_{-39}$	& $-55^{+35}_{-42}$	& $8^{+49}_{-49}$	\\
		time	\T\B	& F	& $21^{+48}_{-51}$	& $8^{+43}_{-42}$	& $-25^{+45}_{-47}$	\\
		polar \T\B	& F	& $-5^{+42}_{-41}$	& $35^{+41}_{-42}$	& $30^{+43}_{-48}$	\\
		\hline\hline
	\end{tabular}
%	\caption{\small %===============here
\tablecomments{
	Amplitude of power spectra $\ell(\ell+1)C^{YY'}_\ell/(2\pi)$
	estimated from difference maps that were constructed with different divisions of the data
	(column `Test').
	Division was done either in time or by polarimeter
	(labeled `time' or `polar' respectively, see text).
	Results are given both for the Bayesian (B) and frequentist (F)	approaches. 
	Error values indicate $68\%$ confidence intervals.
	}
	\label{tab-diff}
\end{table}

%=======================================================different squares (1.1deg, 1.7deg)
\subsubsection{Regions of different sizes}
We also investigated the dependence of the frequentist results on
the size of the square patch chosen for the power spectrum estimation.
The square regions of different size that we used are indicated by the boxes in Figure~\ref{fig-scan}.
The square region of width $1.7^\circ$ is centered at $l=113.23^\circ$, $b=40.2^\circ$.
The square region of width $1.1^\circ$ is centered at $l=113.13^\circ$, $b=40.25^\circ$.
The results are summarized in Table~\ref{tab-squares} and are consistent with the earlier results.
There is no significant increase in the error bars when using a smaller region of the maps
because the edges of the square regions are noisier than the central portion (see Fig.~\ref{fig-scan})
and pixels near the edges have negligible statistical weight in the power spectrum estimation.

%--------------------------------------------Table 5: ps for 1.1 & 1.7 deg^2 ===========here
\begin{table}
	\centering
	\caption{power spectra of different sky sizes} %===========here
	\begin{tabular}{c | ccc}
		\hline\hline
		$\begin{array}{c}
			\textrm{Region Size}\\
			(x^\circ\times x^\circ)
		\end{array}$ & $EE$	& $EB$	& $BB$	\\
		\hline
		$x=1.7$	& $69^{+47}_{-48}$\T\B	& $12^{+52}_{-47}$	& $22^{+50}_{-47}$	\\
		$x=1.1$	& $63^{+48}_{-44}$\T\B	& $8^{+50}_{-42}$	& $14^{+51}_{-44}$	\\
		\hline\hline
	\end{tabular}
%	\caption{\small   %===============here
	\tablecomments{
		Amplitude of power spectra $\ell(\ell+1) C^{YY'}_\ell / (2\pi)$
		estimated from square regions of different size (see Fig.~\ref{fig-scan}).
		The error numbers indicate the $68\%$ confidence intervals.
		}
	\label{tab-squares}
\end{table}

%======================================================= Gaussianity Tests
\subsubsection{Gaussianity test for the maps}

Gaussianity in the pixel-domain signal is an essential assumption 
for the methods of power spectrum estimation that we used.
To test our $Q$ and $U$ maps we applied the Kolmogorov test
to the eigenvalue-normalized Karhunen-Loeve coefficients,
as performed in \citet{ma-wu-ng}.
If the signal is Gaussian, then the K-L coefficients should be normally distributed.
In the process, we found that
some of the eigenvalues of the noise-whitened signal matrix were negative owing to
the high noise and imperfectly estimated signal in those modes. 
We thus excluded these modes from the test, but included all the other modes.
These coefficients passed with a clear margin the Kolmogorov test for Gaussianity 
at 95\% confidence.

%==================================================Beam asymmetry; Q/U leakage
\subsubsection{Beam asymmetry and polarization leakage}

In certain circumstances an asymmetry in the beam may induce spurious polarization signals.
For example, if an asymmetric beam rotates simultaneously with the HWP, 
the resulting $EE$ or $BB$ spectrum will contain power leakage from the $TT$ mode.

Scans of Jupiter were used to quantify leakage from $T$ to $Q$ and $U$. 
Jupiter has an inherent polarization of less than 0.2\% at 140~GHz \citep{millipol}, 
which is small compared to the noise on $Q$ and $U$ during beam mapping.
Out of 12 polarimeters only two showed an instrumental polarization signal at a level of 4\% and 5\%. 
No other polarimeter showed leakage from temperature to $Q$ or $U$ at a level larger than about 1\%,
which was the typical noise level for this measurement.

To quantify this effect on the power spectrum 
we performed an end-to-end simulation.  
Taking the $\Lambda$CDM as the underlying model for $T$ 
we conservatively assumed $3\%$ leakage into each of $Q$ and $U$, 
which is equivalent to 4.2\% instrumental polarization, for all 12 polarimeters.  
The TOD were constructed using the beam patterns as measured in flight.
We processed this signal-only TOD to obtain maps and power spectra.
A Monte-Carlo simulation showed that out of 1,000 realizations 
the {\it largest} contribution of this leakage into the final $EE$ or $BB$ spectrum was 
3~$\mu\mbox{K}^2$ 
for our main bin $\ell=151$--$693$; the mean leakage was smaller.
This test demonstrates that the final results were not affected 
by asymmetries in the measured beam profiles and by the polarization leakage.

%%%%%%%%%%%%%%%%%%%%%%%%%%%%%%%%%%%%%%%%%%%%%%%%%%%
%%%%%%%%%%%%%%%%%%%%%%%%%%%%%%%%%%%%%%%%%%%%%%%%%%%
%%%%%%%%%%%%%%%%%%%%%%%%%%%%%%%%%%%%%%%%%%%%%%%%%%%
\section{Conclusions}
\label{conclusion}

We discussed the analysis of CMB data that were taken with a HWP polarimeter.
Demodulation of the time-domain data based on the rotational position of the HWP 
gave the $Q$ and $U$ data.
These data showed a white-noise spectrum at frequencies well below 50~mHz
for the majority of the data at a level consistent with detector noise.
Most of the data were also Gaussian and stationary. 
We made $Q$ and $U$ maps using a maximum-likelihood technique. 
The maps were also shown to be Gaussian and there was no visible evidence for systematic errors.

We calculated $EE$, $EB$, and $BB$ power spectra using both Bayesian and frequentist techniques.
The Bayesian results gave weak evidence for $EE$ power 
that is consistent with $\Lambda$CDM cosmology and with previous results. 
There was no detectable signal for the $EB$ and $BB$ spectra. 
Results from the frequentist analysis were consistent with the Bayesian ones.
We calculated results for different shape functions and with different priors 
and
found that the significance of detection of $EE$ power was between 83\% and 98\% 
with most of the results giving a probability larger than 90\%. 
We gave results with and without marginalization over calibration uncertainty. 
Inclusion of the calibration uncertainty 
does not change the significance of detection of $EE$ power.

We presented results from tests for systematic errors 
including differencing maps, processing sky regions of different sizes,
assessing Gaussianity, investigating beam asymmetry 
and searching for polarization leakage.  
None of the tests showed evidence for systematic errors.

MAXIPOL is the first experiment to produce CMB data using a modulating HWP. 
The techniques we developed to analyze such data 
should have broad applicability for future CMB experiments 
that are planning to use similar modulation techniques.

%%%%%%%%%%%%%%%%%%%%%%%%%%%%%%%%%%%%%%%%%%%%%%%%%%%
%%%%%%%%%%%%%%%%%%%%%%%%%%%%%%%%%%%%%%%%%%%%%%%%%%%
%%%%%%%%%%%%%%%%%%%%%%%%%%%%%%%%%%%%%%%%%%%%%%%%%%%
\acknowledgments

We thank Danny Ball and the other staff members at the NASA National
Scientific Ballooning Facility in Ft.\ Sumner, New Mexico for their
outstanding support of the MAXIPOL program. 
MAXIPOL is supported by NASA Grants NAG5-12718 and NAG5-3941;
National Science Council, National Center for Theoretical Science, and
Center for Theoretical Sciences, National Taiwan University for J.H.P.~Wu;
PPARC for A.\ H.\ Jaffe and J.\ Zuntz;
a NASA GSRP Fellowship, an NSF IRFP and a PPARC Postdoctoral Fellowship for B.R.\ Johnson; 
the Miller Institute at the University of California, Berkeley for H.~Tran.

We are grateful for computing support 
from the Minnesota Supercomputing Institute at the University of Minnesota, 
from the National Energy Research Scientific Computing Center (NERSC) 
at the Lawrence Berkeley National Laboratory,
which is supported by the Office of Science of 
the U.S.\ Department of Energy under Contract No.\ DE-AC03-76SF00098,
and
from the National Center for High-Performance Computing, Taiwan.
We are grateful for discussions with P.\ Ferreira and members of his research team.
We gratefully acknowledge contributions to the MAXIMA payload by 
A.\ Boscaleri, P.\ de Bernardis, V.\ Hristov,
A.E.\ Lange, P.\ Mauskopf, B.\ Netterfield,
and E.\ Pascale, which were useful for MAXIPOL.

%%%%%%%%%%%%%%%%%%%%%%%%%%%%%%%%%%%%%%%%%%%%%%%%%%%
%%%%%%%%%%%%%%%%%%%%%%%%%%%%%%%%%%%%%%%%%%%%%%%%%%%
%%%%%%%%%%%%%%%%%%%%%%%%%%%%%%%%%%%%%%%%%%%%%%%%%%%

\end{document}